# TreeP: A Tree Based P2P Network Architecture

Benoit Hudzia, M-Tahar Kechadi, and Adrian Ottewill
Department of computer science,
University College Dublin, Belfield, Dublin 4, Ireland.
{benoit.hudzia, tahar.kechadi}@ucd.ie

**Abstract**:  In this paper we proposed a hierarchical P2P network based on a dynamic partitioning on a 1-D space. This hierarchy is created and maintained dynamically and provides a grid middleware (like DGET) a P2P basic functionality for resource discovery and load-balancing. This network architecture is called TreeP (Tree based P2P network architecture) and is based on a tessellation of a 1-D space. We show that this topology exploits in an efficient way the heterogeneity feature of the network while limiting the overhead introduced by the overlay maintenance. Experimental results show that this topology is highly resilient to a large number of network failures.

## I. Introduction

Recently Peer-to-Peer (P2P) systems have emerged as a new distributed computing paradigm. Three major models have rapidly emerged: centralised, decentralised structured and decentralised unstructured [1]. The centralised model, such as Napster [2] and Bittorent [3], uses a central server to index the content and the peers. Usually the P2P systems using this model have the ability to add new servers to sustain their growth. However, they are more vulnerable to attacks such as denial of service by concentrating information in a few selected repositories.

Decentralized structured P2P systems have the advantage of eliminating the need for central servers and provide more autonomy to the participating users. The systems like Gnutella [2], Kazaa [4], and ED2k [5] are the most popular P2P systems in use. However, they rely on a blind flood lookup algorithm [6], or depth first search [7], which are techniques that do not scale well. Several other structured decentralised P2P architectures have been developed and implemented. These include Chord [8], CAN [9], Pastry [10], Tapestry [11], P-Grid [12], and discrete-continuous [13]. All these systems, except for Tapestry, suffer from poor load-balancing capabilities. Some systems implement different strategies in order to overcome this problem, such as routing via nodes with low network latency and/or using super peers to create a highway for communication. However, because they do not scale well, there is still a very high risk of saturating the network and overloading the peers.

Both P2P and Grid have experienced a rapid evolution and widespread deployment. The two technologies appear to share the same objective, which is the pooling and coordination of large sets of distributed resources [15]. In the last few years we have seen various projects that try to combine the complementary features of these two technologies such as NaradaBrokering [16].

In addition, various modifications to the Globus toolkit [17] have been proposed to include P2P technology and thus improving its discovery system [18].

In this paper we propose a hierarchical P2P network architecture based on a dynamic partitioning of a 1-D space. This hierarchy is created and maintained dynamically and provides the DGET [19] grid middleware a P2P basic functionality for discovery and load-balancing. This hierarchy is called TreeP and is based on a tessellation of 1-D space at each level and it exhibits similarities with B+tree. The TreeP topology efficiently uses the heterogeneous aspect of the network while limiting the overhead introduced by the overlay maintenance. We develop an algorithm for creating and maintaining the hierarchy, and then, implement the lookup algorithm on TreeP. TreeP topology is tested for both its robustness and its basic lookup capabilities. Experimental results show that the topology is highly resilient in case of large number of network failures. In the next sections, we will show that it can be easily modified to provide Distributed Hash Table (DHT) functionality. TreeP will be used in the DGET system to provide the basic substrate for the ontology/social P2P search system that is actually under development.

## II.     Related Work

Recently different hierarchy topologies were proposed to work with Distributed Hash Table (DHT). Brocade [22] proposed to organise peers in a two-level overlay, using well-connected super nodes to route messages among different nodes of the network. Garces-Erice *et al.* [23] proposed a hierarchical two-tier topology, which is similar to the Internet hierarchy. Each group of peers is independent and has its internal lookup system. In [24], the authors proposed a topology aware version of Pastry. The next hop is chosen based on the network delay. But the delay is mainly determined by the longest hop, which is, in the case of Pastry, most of the time the last hop. Liebeherr [25] defined a two-level architecture for many-to-many communications in Hypercast based on a Voronoi network [26]. This topology is proven to be scalable and easily maintainable. However, all these proposed architectures, composed of two-layer overlay, do not take into account the load of the super peers and the distribution of nodes / requests over the network.

Srivats *et al.* presented in [26] a multi-tier capacity aware architecture. This system creates groups of peers based on the bandwidth and the resources provided by the peers. Such groups create a layered topology where the most reliable and resourceful nodes are grouped in the top layer while the most unreliable nodes are placed in the bottom layer. The algorithms, used for creating and maintaining an overlay topology, were presented by Beal in [27,29]. Beal introduced a parentless distributed hierarchy formation. He proved that, by using such hierarchical clustering and neighbours relation, one could restrict the failure (a node or a region) as long as it is a convex failure [28].

The use of a tree structure and more specifically B-tree structure [30] has been proposed in many papers. This technique was used in [20] for creating and maintaining a B+tree for distributed indexing purposes. Those approaches aim to add more functionality to a P2P network using existing topology to maintain multiple B-tree mainly for discovery service purposes. The TreeP network architecture rearranges the P2P network overlay in a tree like topology. This allows us to directly leverage the system services based on the network topology characteristics instead of adding another layer of structure/abstractions to the middleware system.

## III. The TreeP Architecture

The overlay topology consists of several layers of peers. Each layer maintains information about its peers. The main objective of this overlay topology is to leverage the capabilities of a heterogeneous network of computers by using a hierarchical structure. Like the architecture proposed in [26], the top layers are composed of more stable and efficient peers and the bottom layers consist of unstable and unreliable peers. The concept of hierarchy based on tessellations of space provides a powerful organisational framework. In other words, the tessellation of the network provides a generic basis for a broad class of spatially constrained hierarchical and fractal networks. The algorithms used in tessellations can also be adapted to take into account the locality of peers with low overhead.

The main objectives of the TreeP architecture are twofold: 1) Provide users with a network with a minimum maintenance overhead and self-adaptable; 2) Provide users with a network that is able to evolve with the system fluctuations by tailoring the granularity of the overlay to respond to various flows.

We first assume that each node in the system has a unique identification number (ID). This ID can be self-generated by the system or by an external ID system (e.g., certificate authority). The system nodes do not necessary need to use the same ID each time they connect to the network. However, using the same ID for subsequent connections can be useful. For instance, a user might want to keep the same ID so it can be found easily. The IDs can be assigned randomly or based on a hash of the IP/Port numbers. We can even use a system that provides localisation capability to TreeP similar to the anchor system in [34]. Other scenarios can invoke a preliminary search for an ID range to choose from. This will allow the system to maintain a balanced tree. In the TreeP architecture the ID provides a spatial coordinates in the system. These will be mapped onto a 1-D space, providing a virtual localisation of the peers. Using this localisation, we will create a tessellation of the corresponding space.

The TreeP P2P network is a UDP based overlay architecture, as UDP/IP doesn't need to handle continuous streams of data between peers for creating, maintaining, and using the network. As DHTs are designed without hierarchical routing capabilities, all TreeP peers are equal and use the same rules for determining the routing paths for lookup messages. This approach is different from the hierarchical routing on the Internet. In TreeP the network of nodes create a tree or hierarchical topology. We have used the tree structure as it has been widely employed and has been proven to be very efficient for resource discovery and other services [31].

### a. The TreeP Network Topology

The TreeP topology consists of connections (actively maintained) that provide the skeleton of the hierarchy. The routing table depends mainly on these connections. The hierarchy is created starting from the bottom layers. In other words, the joining peers are assigned to the lowest and are promoted to upper layers to fit the needs of the systems (load balancing, replacement of a leaving node, etc.). The systems promote nodes in a distributed manner and the criteria used for promotion are based on the characteristics of the nodes such as: CPU, Memory, Bandwidth, network load, systems load, Uptime and Storage Space.

The network is assumed to be spatially distributed. The space is divided into regions and each region is a set of locations (IDs). Each region is associated with one level of the hierarchy. Without loss of generality we assume that the ID of a node $v_{ij}$ provides a spatial location in its region. Let *d(u, v)* be the Euclidian distance between the nodes *u* and *v* and let $P_u$ be the position of the node *u*. The network can be represented by an undirected graph *G=(V, E)* where *V* is a set of vertices and *E* is a set of edges (virtual links). The degree of a node *u* represents the number of nodes connected to it and it is denoted by $deg_u$. The region $\Theta_0$, corresponding to the level 0, contains all active nodes of the network. In the initial network $\Theta_0$, the connections are established according to the needs of the active nodes. Each node needs to maintain a minimum of two connections.

The TreeP structure matches the nodes needs perfectly and avoids maintaining unnecessary edges. With the TreeP network topology, communications between two nodes ensure that both nodes are alive, thus avoiding costly maintenance and keeping control messages to a minimum. If they stop interacting and have more than two edges, each node can safely delete the other from their routing table. Such modification is purely local to each node and doesn't affect the rest of the network.

The levels $\Theta_j$ (j > 0) consists of nodes that have been selected from level 0. These nodes have been promulgated by their parents or during the election process (a node is elected if it reaches a degree of 2 and does not have a parent). In addition, these nodes have been chosen based on their uptime and their resource characteristics. Unlike $\Theta_0$, each $\Theta_j$ (j > 0) has a bus topology linking all its nodes together. Therefore, each node has two neighbours except the two endpoints and is responsible for its tessellation (see Figure 1).

Each node at level *k* is a parent of the nodes covered by its tessellation at level *k-1* (level that belongs to its $\Theta_j$ 1-D space as shown in Figure 1). A parent has a maximum number of nodes that can maintain. This maximum is either defined at start up or can be dynamically calculated using the nodes' characteristics and their actual load. A parent is also responsible for promoting a child to its level of the hierarchy.

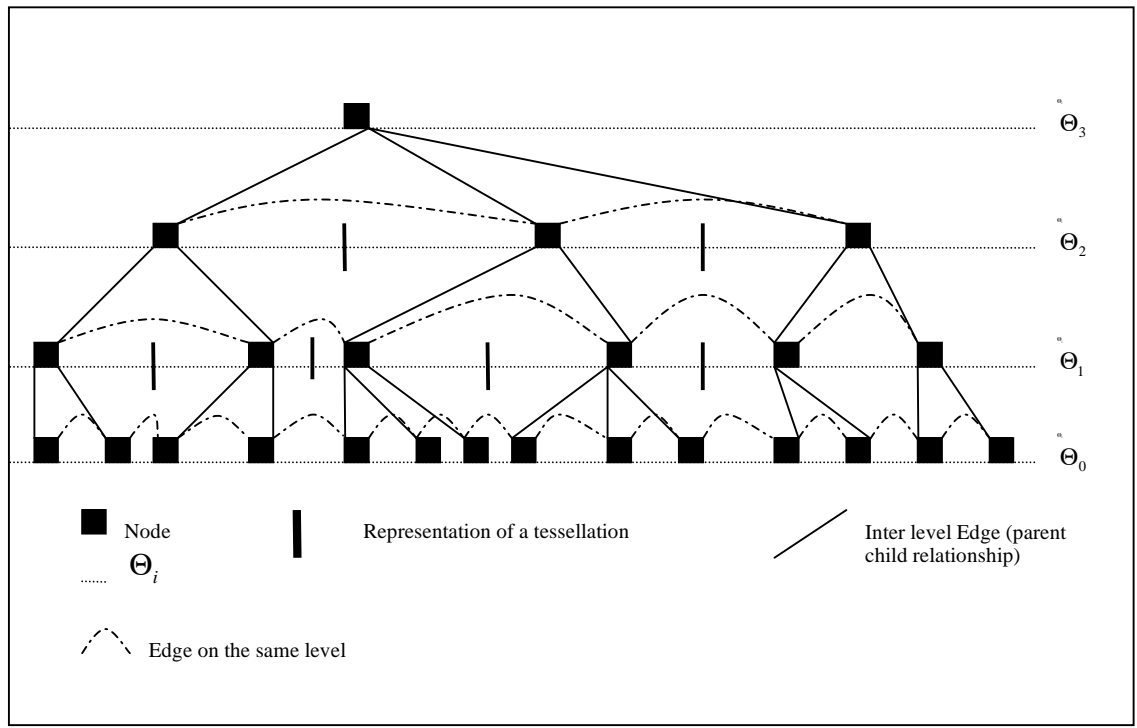

| ■ Node | | Representation of a tessellation | | Inter level Edge (parent child relationship) |
| ...... $\Theta_i$ | | | | |
| ⌒ Edge on the same level | | | | |

**Figure 1: TreeP Skeleton**

All the edges of the hierarchy (called active connections) are actively maintained. The type of maintenance needed may differ from one edge to another. A parent doesn't explicitly maintain the connection with its children. If they do not report regularly they will be simply be deleted from its routing table. Connections within the same level are maintained by all the nodes of that level.

The TreeP structure is similar to a B+tree. However, unlike B+tree the nodes of level $\Theta_0$ can also be part of any other level. Another main difference is that the nodes within the same level are connected by a bus topology, hence, avoiding unnecessary communications through other levels.

### b. TreeP Creation

The creation and maintenance of a TreeP structure is quite simple. When a node reaches a degree of 2 and does not have a parent, it will search for a parent by contacting its neighbours. The election of a parent is triggered when a node reaches a degree of 2. The election technique used in TreeP is described in [28]. When the election is triggered, each participating node starts a countdown. The initial value of the countdown is calculated according to the node characteristics (CPU, bandwidth, average work-load, average network load, etc.). A node that has higher characteristics will have smaller countdown initial value. When the countdown of a node reaches 0 and if no other node was elected during this time, it will signal to its neighbours that it is their new parent. Similarly, if a parent has less than two children, it will start a countdown, but this time, the higher is the characteristic the longer is the countdown. At the end of the countdown, if it still has less than two children it will leave its current level and will become an ordinary node of the level 0.

### c. The routing table

Each peer maintains its routing table by exchanging data through its active connections. The main information stored in the routing table is a set of tuples (ID, IP, Port). The routing table system is composed of six tables:

- **Level 0 routing table**: every node has one since all the nodes belong to level 0.
- **Level *i*, (*i* > 0), routing table**: concerns the nodes of the same level. It contains direct and indirect neighbours (direct neighbours of direct neighbours) list in that level. In addition, it contains information about the parents of level *i* of its direct neighbours at level 0. It also contains information on the direct neighbours of level 0 that belongs to the same level i.
- **Children routing table**: only nodes that belong to a Level *i* with i > 0 maintain this table. It contains a list of nodes that belong to the tessellation of the node that is responsible for it. It also contains a list of children of its direct neighbours.
- **Level 1 parent**: all nodes have a parent node.
- **Superior Node List**: consists of nodes with more than one level. Basically, it contains a list of its ancestors (red nodes in Figure 2) and direct neighbours of its immediate parent. This replication of information provides a higher degree of robustness at minimum cost.

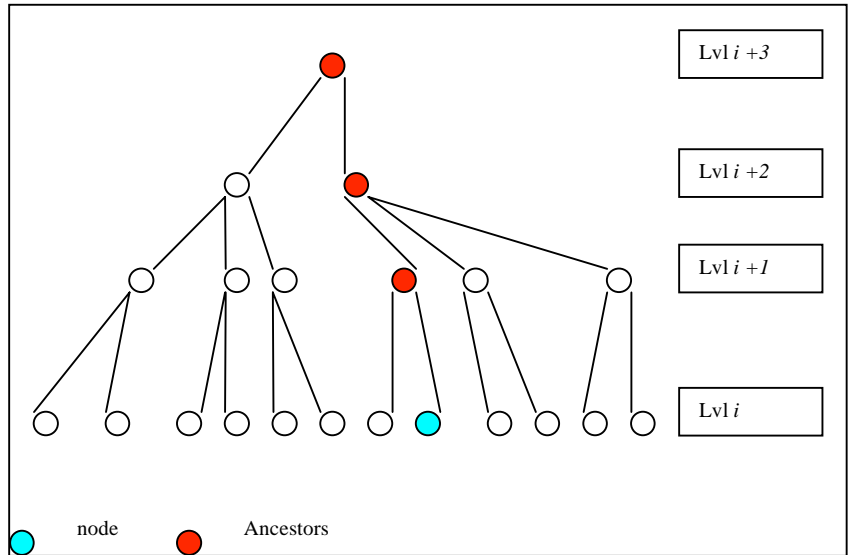

**Figure 2: Ancestors**

All the entries in the routing table have a timestamp associated with each node providing the information. This timestamp is reset at every occurrence of an active communication with the corresponding node. The entry will be deleted after the expiration of the timestamp.

### d. Maintaining the routing table

Unlike some systems such as Chord [8], the TreeP routing table is maintained in a very efficient way in order to minimize the data exchange between the nodes. The exchange concerns only the routing table information that is out-of-date. When two nodes communicate for the first time they exchange information about their resources and state: hardware, network capacity, current CPU load, network load, etc. Then, each node has to maintain this information with its direct neighbours. If the connection between two nodes *a* and *b* is at level 0 and they have different maximum levels $i_a$ and $i_b$, (with $i_a < i_b$), then *a* will send to *b* information about its parents at the level $i_b$.

Each active connection at level *i*, (*i > 0*), allows the two end points to exchange their inner neighbour's entry information and also information about their children. They also exchange their routing table entry about their immediate parent of level *i+1*. If the parent entry doesn't exist it will be added and then forwarded to its own parent. Such exchange prevents the network from having two roots of the tree that are not connected. Finally for any two neighbours, after the initial synchronisation and the usual keep-alive message, they only exchange information concerning the out of dated data. Sometimes, the update can be delayed, waiting to be piggybacked during a "keep-alive" exchange. In the current implementation the update is exchanged immediately. This technique, for maintaining the routing tables, provides better connectivity and, therefore better performance and fault tolerance.

### e. Characteristic of the architecture

TreeP shares many common features with B-trees. For instance the height $h$ of the hierarchy can be calculated in the same way as for a B-tree [33]. The height of a TreeP network having $n$ nodes and a minimum degree $t$, ($t \geq 2$), is given by: $h \leq \log_t((n+1)/2)$. Usually, the height of a tree corresponds to the average number of children per parent $c$: $h = \log_c((n+1)/2)$.

In order to determine the size of the routing table the following definitions are required: Let $c$ be the average number of children per node in a TreeP network of $n$ nodes. Let $l_i$ be the number of connections at level $i$, and let $c_a$ be the number of children of a node $a$. $c_i$ is the number of children of the direct neighbours at the same level. $d_a$ is the number of direct neighbours of a node $a$. $L_i$ is the number of node $i$ entries gathered from level 0 connections to its children. $n_i$ is maximum number of nodes in level $i$. Based on previous architecture description, we have: $2 \leq l_0 \leq n-1$, $0 \leq l_i \leq 2$, $0 \leq d_a \leq 2$, and $0 \leq L_i \leq n_i$. So, the size of each routing table is as follows:

- For nodes belonging only to level 0, the size is given by  $(l_0 + h)$.
- For nodes that are in level $i$, $i > 0$, this is given by $(l_0 + l_i + L_i + c_i + c_a + d_a + h - i)$.

As we can see a node of level 0 (the majority of the nodes in the network) has a very small routing table and has to maintain only $(l_0 + 1)$ active connections. The nodes of level 1 have to maintain $(l_0 + c_a + d_a)$ connections. Finally, nodes in upper layers (i > 1) will maintain $(l_0 + c_a + d_a + 2)$ active connections. The number of actively maintained connections per node is reasonably small and this demonstrates that the topology makes efficient use of the heterogeneity of the network.

### f. Routing/Lookup algorithm

The procedure used for lookup and routing is based on a distance measure function $D(a, b)$ defined as follows:

$$\begin{cases} lvl_a = 0 \rightarrow D(a,b) = d(a,b) \\ D(a,b) = 0 \rightarrow (d(a,b) - \dfrac{L}{2^{(h-lvl_a)}}) \leq 0 \\ D(a,b) = d(a,b) - \dfrac{L}{2^{(h-lvl_a)}} \end{cases}$$

Where $a$ and $b$ are two nodes and $lvl_a$ is the maximum level of a node $a$.

This distance function $D(a, b)$ takes into account the location of $a$ and $b$ in the topology and the size of their tessellations. A greedy algorithm for lookup and routing is given in Figure 3. In addition of the greedy algorithm we propose two other different algorithms:

- Non-Greedy (NG): the algorithm returns a node $n$ that verifies the condition; $d(a, n) - d(a,x) < 0$. The procedure basically ends when a node satisfying the condition is found.

- Non-Greedy With Fall Back (NGSA): the algorithm selects a node *n* such that *d(a, n) – d(a, x) < 0*. In addition, other nodes satisfying the criteria are also selected. This algorithm provides alternative routing paths in the case that the maximum TTL allowed is reached or the current path is a dead end. These additional routing paths are provided at the expense of adding data to the request.

The complexity of these three routing algorithms is *O(log n)* due to the TreeP topology

---

A is the address of the Current Node, N is the Next Hop, X is the ID of the node/Object to resolve.
Level_*I* is the maximum level a Node belongs to.

IF TTL > 255 THEN discard the request
IF target X is in the routing table THEN transmit back the result
ELSE IF request from the parent of Level 1
    N = Search_Level_Zero()
    ELSE N = Search_level_A()
    IF N != A
    IF Request From parent of Level1 THEN forward the Request to N
    ELSE IF $D(n,x) \leq D(a,x) \times \frac{1}{2}$ THEN forward the request to N
    ELSE IF Level_A == 0 Then forward the request to N
    ELSE IF query from parent of Level_L Where Level_L = Level_A +1
        THEN forward to N
    ELSE IF Superior_Node_List_Not_empty()
        THEN forward the request to the Node that is the closest to X satisfying $D(n,x) \leq D(a,x) \times \frac{1}{2}$, IF none match the criteria THEN send the request to the superior node with the highest level
        ELSE Reply Not Found
    ELSE IF Request from parent of level 1 THEN Reply Not Found
        ELSE IF Level_A == 0 THEN N = Closest_Child(X)
        Send to N
    ELSE IF Superior_Node_List_Not_empty()
        THEN forward the request to the Node that is the closest to X satisfying $D(n,x) \leq D(a,x) \times \frac{1}{2}$, IF none match the criteria send the request to the superior node with the highest level
        ELSE Reply Not Found

**Figure 3: Routing algorithm: G**

The greedy routing algorithm (G) comes with two main problems: firstly, it does not always find a path between two nodes, (as shown in Figure 4). Secondly, it is not loop free in case of heavily disconnected topology. As the nodes can join and leave at any time, these dynamic operations may leave parts of the network disconnected and isolated from the rest. This will cause the greedy algorithm indefinitely looping when the source and destination nodes are located in two disconnected parts of the network.

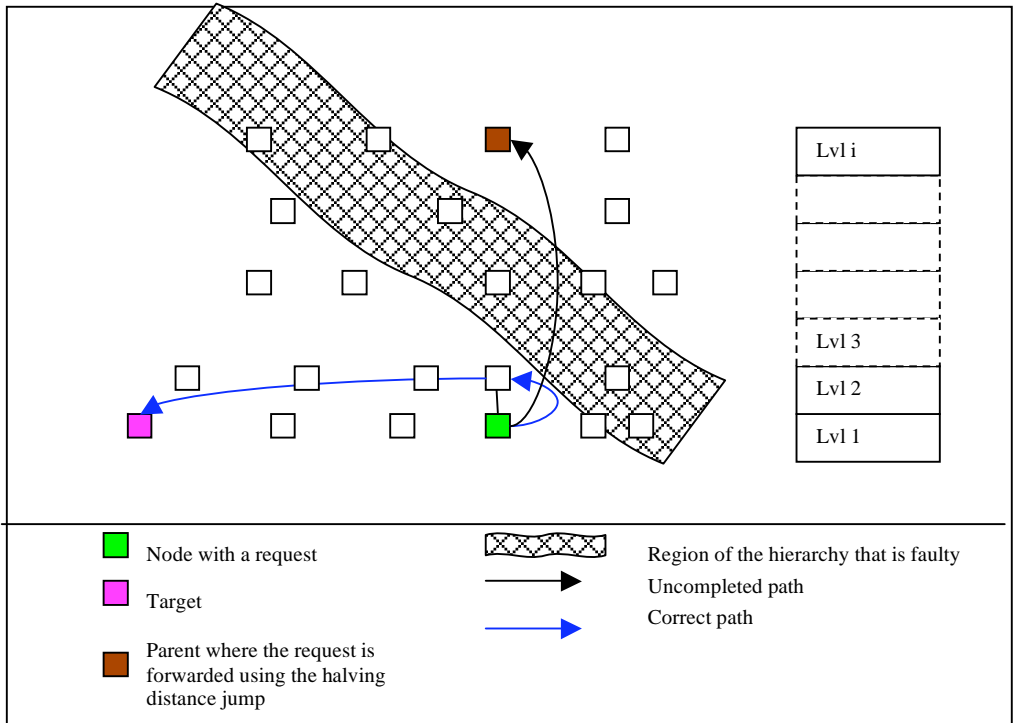

**Figure 4: Non loop free algorithm**

To prevent such problem from occurring, a TTL counter is used. This counter is also useful for finding a routing path to the destination. When a node receives a request and it has a TTL greater than the height of the hierarchy, the Euclidian distance is used instead of using the distance $D(x, y)$. This provides a finer grained routing, because a request that has a higher TTL means that the network is unstable and/or disrupted. In addition, this augmented routing algorithm also supports a search for an object associated with ID, which can be used for resource discovery.

## IV. Experimental Results

In this section the TreeP performance and robustness are evaluated using the 3 different lookup algorithms. These algorithms have direct impact on TreeP performance and on the load of the network nodes.

The implementation is based on a packet switching protocol. Each peer has its own routing table, therefore the routing decisions are made locally to each node without knowledge of the global state of the network. In order to evaluate the TreeP architecture two different cases are considered: In the first case the maximum number of children (*nc*) is fixed to 4 while in the second *nc* is defined according to the nodes capabilities such as CPU, Memory, bandwidth, etc. Note that the initial TreeP network is randomly chosen.

In this experimental evaluation we observe the efficiency of the different lookup algorithms and their impact on the TreeP network. The TreeP network topology is dynamic since the node can leave and join at any instant in time. The performance evaluation is done when the system reaches its steady state, which is based on the maximum hierarchy size. Note that for each case described above the TreeP topology is different and depends of the constraints imposed on the system. The three routing algorithm are tested during the steady state. We randomly disconnected some nodes at a rate of 5% and observed the behaviour of these routing algorithms, until the number of the remaining nodes reached a threshold of 5% of the initial topology. This technique of testing puts a huge stress on the routing algorithms, but at the same time it allows a reasonably good evaluation of the TreeP robustness.

### a. First case, $nc = 4$, $h = 6$

As we can see in Figure A of the appendix all the algorithms are robust against random disruption of the network at its steady state. When 30% of the nodes have been removed 10% of the lookups have failed. When 50% of the nodes are disconnected, 25 to 30% of the requests are lost. We can see these algorithms achieve similar performance with a fluctuation of 2%. The NGSA algorithm is not performing much better than the NG or the Greedy algorithm, which is a surprising result. The gain obtained by the NGSA algorithm compared to its cost in terms of bandwidth makes it less attractive to be used with this topology.

Figure B shows that the average number of hops to reach the destination is independent of the rate of failed nodes. Above 70% of disconnected nodes, the TreeP network is composed mainly of several isolated sub-networks.

Figure E shows the maximum and minimum of failed lookups. One can notice that when 35% of the nodes are disconnected the maximum number of failed hops increases dramatically. This is due to the fact that the network is partitioned into two different isolated sub-networks.

Figures F and G present the percentage of requests (z axis, 0:50) that have been resolved in a certain numbers of hops (y axis, 0:30) depending on the percentage of failed nodes (x axis, 0:80). The NGSA algorithm graph is not presented as it was found to be almost identical to the NG algorithm graph. We can see that the number of hops is constant (5 hops average). Such behaviour is very good and demonstrates that the routing technique is stable and efficient. The main difference between NG, NGSA, and Greedy is that the greedy algorithm has a higher rate of requests that have been resolved in less than 5 hops: 4 hops represent ~50% for G and represent ~45% for NG.

### b. Second case, h=6, $nc$= variable

With a variable maximum number of children, we can see in Figures H and I that the curves are much steeper. Both curves have a peak for 5 hops representing 60% of the requests in the system. The TreeP performance is really affected when 40% of the nodes are disconnected as in the previous experiment. Figure C also shows that the behaviour of the algorithms is similar to the first case (experiment).

Compared with a fixed maximum number of children, the average number of hops depends with the number of nodes that have been removed (see Figure D). This variation in the average number of hops becomes more important when more than 30% of the nodes have left the system. This shows that having a variable number of children in TreeP only affects the average number of hops to resolve the IDs when there is a large number of failures. The unbalanced hierarchy remains effective.

Figure D also shows that there are few differences between the two approaches. We can notice that the number of hops increases, but the cost of this increase is easily compensated by the fact that TreeP is based on a variable maximum number of children, which provides an efficient use of the heterogeneity of the network. We can also have a flattened TreeP hierarchy, which greatly reduces the number of hops per request, provides better bandwidth consumption and has a more stable topology.

## V. Conclusion

In this paper we presented a general framework for creating robust and efficient use of the heterogeneity of the hierarchical P2P systems. The TreeP topology is self-maintained and self-healing, with low maintenance cost and resilience to failures. The TreeP topology allows the users to take advantage of the different peers' characteristics and the ability to rapidly adapt to different situations (load balancing, failures, network traffic, etc.). TreeP is well suited to heterogeneous P2P networks. Our experimental results show that TreeP efficiently handle failures. It is highly stable and the resource oriented-hierarchy provides an optimal use of the network. The technique used to define and maintain the routing tables allows the use of routing algorithms having a logarithmic complexity. We have shown that the use of adaptive algorithm for maintaining the TreeP topology improves the overall performance and provides low overhead in case of large number of failures. The TreeP architecture shows great potential based on its adaptability and its capacity to react and heal itself in harsh environment.

## VI. Future Works

We are in the process of implementing a library that will be tested on the Grid 5000 test bed. We will stress the TreeP topology in a real-time environment with various churn rates, loads, and heterogeneous resources. We will also adapt the topology to a 2D space (using Voronoi tessellations) to provide a higher degree of reliability and stability.

With regard to the discovery system, we are in the process of exploring the possibility of mapping tree ontology onto the topology and other more classical discovery systems for P2P.

We also need to evaluate the impact of the different strategies for assigning an ID to a node and for electing a parent, which has less than two children at the end of the countdown. An alternative strategy is that if the node is in level $i > 1$, it maintains its current status even if it doesn't have any children. This approach may keep stable and powerful nodes within the upper layers of the hierarchy.

## *VII.* **References***:*

[1] Q. Lv, P. Cao, E. Cohen, K. Li, S. Shenker, Search and Replication in Unstructured Peer-to-Peer Networks, Proc. of 16[th] ACM International Conference on Supercomputing (ICS'02), New York, USA, June 22-26, 2002.

[2] S. Saroiu, K. P. Gummadi and S. D. Gribble, Measuring and analyzing the characteristics of Napster and Gnutella hosts, Journal of Multimedia System, Vol. 8, Issue 5, August 2003.

[3] B. Cohen, Incentives Build Robustness in BitTorrent, in Proc. of 2[nd] Workshop on Economics of Peer-to-Peer Systems, Berkeley, CA, USA, June 5-6, 2003.

[4] N. Leibowitz, M. Ripeanu, A. Wierzbick, Deconstructing the Kazaa Network, 3[rd] IEEE Workshop on Internet Applications (WIAPP'03), San Jose, CA, USA, June 23-24, 2003.

[5] T. Hossfeld, K. Leibnitz, R. Pries, K. Tutschku, nformation Diffusion in eDonkey Filesharing Networks, Australian Telecommunication Networks and Applications Conference, December 8-10, 2004.

[6] J Ritter, "Why Gnutella Can't Scale. No, Really", http://www.darkridge.com /~jpr5/doc/gnutella.html.

[7] B. Krishnamurthy, J. Wang, Y. Xie, Early measurements of a cluster-based architecture for P2P systems, ACM SIGCOMM Internet Measurement Workshop, San Francisco, CA, USA, November 1-2, 2001.

[8] I. Stoica_, R. Morris, D. Karger, M. F. Kaashoek, H. Balakrishnan, Chord: A scalable peer-to-peer lookup service for internet applications, ACM SIGCOMM Internet Measurement Workshop, San Francisco, CA, USA, November 1-2, 2001.

[9] P. Druschel and Rowstron, A. Storage Management and Caching in PAST, a Large-scale, Persistent Peer-to-peer Storage Utility, 18[th] ACM Symposium on Operating Systems Principles (SOSP 2001), Lake Louise, AB, Canada, October 2001.

[10] S. Ratnasamy, P. Francis, M. Handley, R. Karp, A Scalable Content-Addressable Network, SIGCOMM Internet Measurement Workshop, San Francisco, CA, USA, Nov.1-2, 2001.

[11] A. Rowstron, P. Druschel, Pastry: Scalable, distributed object location and routing for large-scale peer-to-peer systems, IFIP/ACM International Conference on Distributed Systems, Heidelberg, Germany, November 12-16, 2001.

[12] B.Y. Zhao, J. Kubiatowicz, AD. Joseph, Tapestry: an infrastructure for fault-tolerant wide-area location and routing, U. C. Berkeley Technical Report UCB//CSD-01, 2001.

[13] K. Aberer, P. Cudre-Mauroux, A. Datta, Z. Despotovic, Grid: A Self-Organizing Access Structure for P2P Information Systems, 9[th] International Conference on Cooperative Information Systems (CoopIS'01), Trento, Italy, September 5-7, 2001.

[14] M. Naor, U. Wieder, Novel architectures for P2P applications: the continuous-discrete approach, 15[th] ACM Symposium on Parallelism in Algorithms and Architectures, (SPAA'03), San Diego, CA, USA, June 7-9, 2003.

[15] I. Foster, A. Iamnitchi,"On Death, Taxes, and the Convergence of Peer-to-Peer and Grid Computing", 2[nd] International Workshop on Peer-to-Peer Systems (IPTPS'03), Berkeley, CA, USA, February, 20-21, 2003.

[16] S. Pallickara, G. Erlebacher, D. Yuen, G. Fox and M. Pierce, NaradaBrokering as Middleware Fabric for Grid-based Remote Visualization Services.," Invited Talk at the American Geophysics Union (AGU) Fall Meeting. San Francisco, CA. December 2003.

[17] I. Foster, C. Kesselman. "Globus: A Metacomputing Infrastructure Toolkit." Intl. J. Supercomputer Applications, 11(2): 115-128, 1997.

[18] D. Talia and P. Trunfio, "Toward a Synergy between P2P and Grids", University of Calabria published in IEEE Internet Computing, 7(4):95-96, July, 2003.

[19] B. Hudzia, L. McDermott, T.N. Ellahi, T. Kechadi, Entity Based Peer to Peer in a Data Grid Environment, 17th IMACS World Congress, Paris, France, July 11-15, 2005.

[20] A. Crainiceanu, P. Linga, J. Gehrke, J. Shanmugasundaram, P-tree: a P2P index for resource discovery applications, 13th International World Wide Web Conference, New York, USA, May 17-22, 2004.

[21] C. Baquero, N. Lopes, B+tree on P2P: Providing content indexing over DHT Overlays, Technical Report, Universidade do Minho, March 2004.

[22] BY. Zhao, Y. Duan, L. Huang, AD. Joseph, J. Kubiatowicz, B rocade: Landmark Routing on Overlay Networks, 1st Int'l. Workshop on Peer-to-Peer Systems (IPTPS'02), Cambridge, MA, USA, March 7-8, 2002.

[23] L. Garcés-Erice, E.W. Biersack, P.A. Felber, K.W. Rossand G. Urvoy-Kelle, Hierarchical Peer-to-Peer Systems, Int'l. Conference on Parallel and Distributed Computing (Euro-Par'03), Klagenfurt, Austria, August 26-29, 2003.

[24] M. Castro et al., Topology-Aware Routing in Structured Peer-to-Peer Overlay Networks, Tech. Rep. MSR-TR-2002-82, Microsoft Research, one Microsoft Way, Redmond, WA, 98052, 2002.

[25] J. Liebeherr, HyperCast: A Protocol for Maintaining Multicast Group Members in a Logical Hypercube Topology, Networked Group Communication, pp: 72-89, 1999.

[26] S. Singhal, M. Zyda, Networked virtual environments: design and implementation, ACM Press/Addison-Wesley Publishing Co., New York, NY, 1999.

[27] M. Srivatsa, B. Gedik, L. Liu, Scaling unstructured peer-to-peer networks with multi-tier capacity-aware overlay topologies, Proc. of the 10th International Conference on Parallel and Distributed systems (ICPADS'04), Newport Beach, CA, USA, July 7-9, 2004.

[28] J. Beal, parentless Distributed Hierarchy Formation, Technical report IA Lab MIT, 2003.

[29] J. Beal, A robust amorphous hierarchy from persistent nodes, Proc. of IASTED Conference on Communication Systems and Networks (CSN'03), Benalmadena, Spain, Sept. 8-10, 2003.

[30] J. Beal, Near-optimal distributed failure circumscription, 15th IASTED Int'l. Conference on Parallel and Distributed Systems (PDCS'03), Marina de Rey, CA, USA, November 3-5, 2003.

[31] D. Comer, ubiquitous B-tree, Computer Survey, Vol 11, No 2, 1979.

[32] Cormen, Leiserson *et al*, Introduction to algorithm, ACM, 1990.

# VIII. Appendix

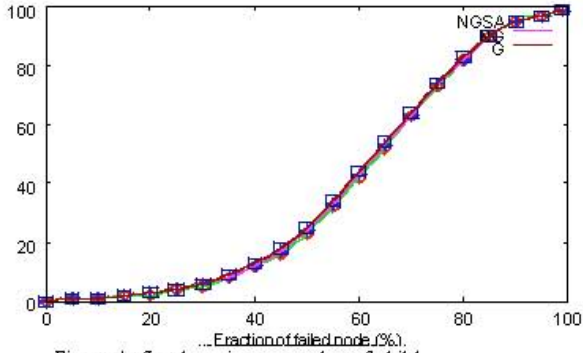
Figure A: fixed maximum number of child

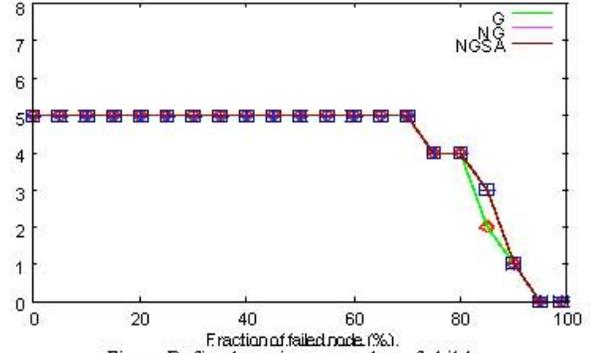
Figure B: fixed maximum number of child

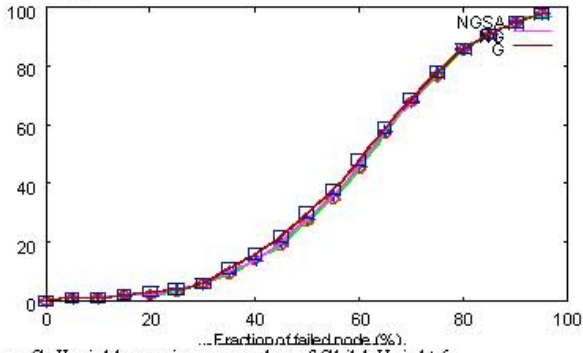
Figure C: Variable maximum number of Child, Height 6

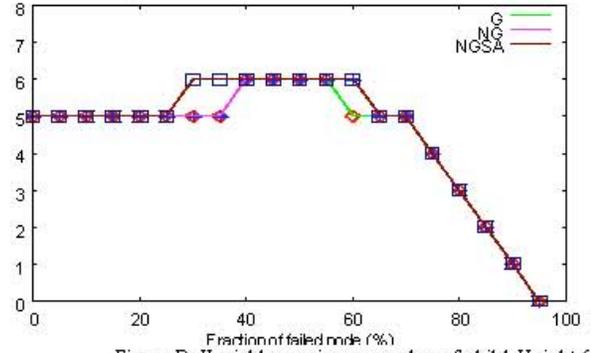
Figure D: Variable maximum number of child, Height 6

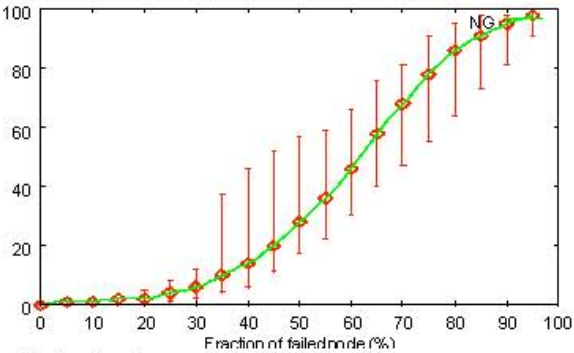
Figure E : fixed maximum number of child
E : fixed

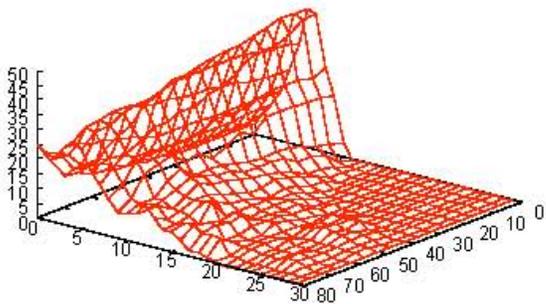
Figure F Greedy : fixed maximum number of child

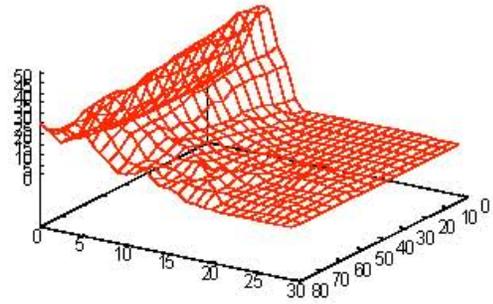
Figure G: NG  fixed maximum number of child

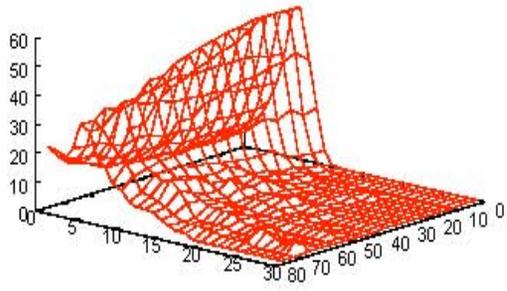 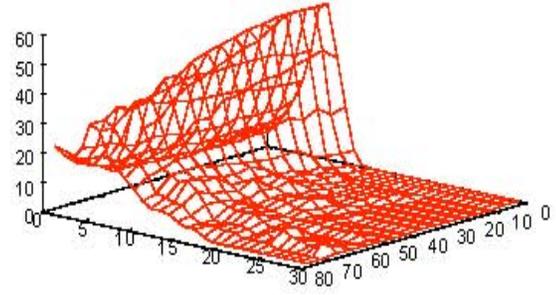

Figure H: Greedy, Variable maximum number of child, Height 6    Figure I: Non Greedy, Variable maximum number of child, Height 6